\documentclass[11pt,a4paper]{article}

\usepackage{jinstpub}
\usepackage{lineno,hyperref}
\usepackage{comment}
\usepackage{color}
\usepackage{siunitx}
\usepackage{textcomp}
\usepackage{capt-of}
\DeclareSIUnit[number-unit-product = ]\percent{\char`\%}
\usepackage[nowatermark]{fixmetodonotes}

\newcommand{\GdSulfate}{Gd$_{2}$(SO$_{4}$)$_{3}$~}
\newcommand{\Hydroxide}{OH$^{-}$~}
\newcommand{\Sulfate}{SO$_{4}^{2-}$~}

\newcommand{\Nitrate}{NO$_{3}^{-}$~}

\title{\boldmath Development of an ion exchange resin for gadolinium-loaded water}

\author[a]{V. Fischer, \note{Corresponding author.}}

\emailAdd{vfischer@ucdavis.edu}

\author[a]{J. He,}

\author[a]{M. Irving,}

\author[a]{R. Svoboda}

\affiliation[a]{Department of Physics, University of California, Davis, CA 95616, USA}


\arxivnumber{2004.04629}

\abstract{
Large water Cherenkov detectors have been successfully used for decades in high- and low-energy particle physics.
Nevertheless, detecting neutrons remains a challenge for such detectors since a neutron capture on a hydrogen atom doesn't release a sufficient amount of gamma energy to be observed efficiently.
The use of gadolinium in the form of soluble salts has been explored extensively to remedy this issue, as gadolinium exhibits both a very large neutron capture cross section and a subsequent high-energy gamma cascade.
However, in order for large gadolinium-loaded detectors to operate stably over long time periods, water optical transparency must be maintained by {\it in situ} purification. New methods have been developed involving band-pass molecular filtering. While these methods are very successful, they are expensive and consume considerable power and space as they seek to minimize loss of gadolinium while removing other impurities. For smaller detectors where some gadolinium loss can be tolerated, a less expensive way to do this is very desirable.  In this paper, we describe the design, development and testing of a system used to purify the gadolinium-loaded water in the 26-ton ANNIE neutrino detector.
}

\begin{document}
\maketitle


\section{Introduction}
\label{sec:intro}

Large water Cherenkov detectors such as IMB~\cite{BeckerSzendy:1992hr}, Super-Kamiokande~\cite{Fukuda:2002uc}, and SNO~\cite{Boger:1999bb} are multi-purpose instruments with rich physics programs ranging from neutrino oscillations searches at the multi-GeV scale~\cite{Fukuda:1998mi} to solar neutrino detection at the MeV scale~\cite{Abe:2016nxk}.
The efficiency of detecting low-energy charged particles in such detectors is however limited by the very nature of the Cherenkov process.
Indeed, in the case of an electron, a kinetic energy above about 800~keV is required to produce Cherenkov photons in water.
The consequence of this threshold is that the kinetic energy of a charged particle isn't fully converted to Cherenkov light.
This effect is accentuated at low energies where a single gamma ray, for example a 2.2~MeV gamma from capture on hydrogen, can undergo several Compton scatters that generate electrons near or below the Cherenkov kinetic energy threshold.
For this reason, pure water Cherenkov detectors are not efficient at detecting neutrons.

To address this issue, the GADZOOKS!~\cite{Beacom:2003nk} program introduced the idea of adding a sub-percent amount of soluble gadolinium (Gd) salts in large water Cherenkov detectors to increase their neutron detection capability.
Gadolinium has the largest known thermal neutron capture cross section of all natural elements ($\sim$49,000~barns)~\cite{ENDF8}, a value overwhelmingly driven by the cross sections of the $^{155}$Gd ($\sim$60,800~barns) and $^{157}$Gd ($\sim$254,000~barns) isotopes with respective abundances of 14.8\% and 15.7\%.
Upon neutron capture, the subsequent energy released in the form of a gamma cascade is equal to 8.54 and 7.94~MeV, respectively, for an average of about 8~MeV.
Due to the gamma multiplicity of the cascade and the Cherenkov threshold of electrons in water, these 8~MeV gamma cascades produce Cherenkov photons roughly equivalent to 5~MeV electrons, enabling a higher detection efficiency. 
This has been verified experimentally by the EGADS demonstrator experiment~\cite{Ikeda:2019pcm}, discussed in more detail below.

Gadolinium salts exist in the form of several compounds, mainly with gadolinium in the +3 oxidation state.
Studies have been performed with gadolinium nitrate, Gd(NO$_{3}$)$_{3}$, gadolinium chloride, GdCl$_{3}$~\cite{Coleman:2008ng}, and gadolinium sulfate, Gd$_{2}$(SO$_{4}$)$_{3}$, but the large light absorption of NO$_{3}^{-}$ in the UV - the dominant region of the Cherenkov spectrum - and the high reactivity of chloride relative to sulfate drove the decision to focus R\&D efforts on gadolinium sulfate.
Central to these efforts is the large scale program called EGADS~\cite{Ikeda:2019pcm}, a 200-ton instrumented water detector built to study the effect of Gd-loaded water on the capabilities of water Cherenkov detectors.
One of the main issue EGADS was designed to address was water purification.

In a pure water Cherenkov detector, water must be kept free of particulates, bacteria, and dissolved UV-absorbing ions and molecules to reach high levels of light transparency. 
Since materials in contact with the water leech such contaminants into the bulk, {\it in situ} purification is required.
Water purification is performed by the use of filters of various pore sizes, sterilizing ultraviolet lamps, reverse osmosis systems, and ion exchange resins, the latter two being the most efficient at removing ionic impurities.
When gadolinium is added to ultra-pure water in the form of soluble salts, it becomes a dissolved ion that would be removed by conventional purification systems. 
EGADS developed a water purification system capable of keeping a high level of light transparency while not removing gadolinium sulfate from solution.
To do so, it utilized an intricate system of ultra-filters, nano-filters, and an anion exchange resin not commercially available and of proprietary composition~\cite{PrivateVagins}.

The use of gadolinium sulfate as a dopant to enhance the neutron sensitivity in water Cherenkov detectors is becoming more wide spread~\cite{Sweany:2011ai, Back:2017kfo, Askins:2015bmb} and thus the interest in water purification is increasing. In this paper we will discuss the development of a small system for the Accelerator Neutrino Neutron Interaction Experiment (ANNIE). 
For ANNIE, the relatively small size of the detector ($\sim$25~tons) and short optical paths ($\sim$ 4~meters) makes the use of the complex state-of-the-art EGADS water purification system unnecessary and expensive.  
Therefore, a smaller and less expensive purification system better suited for the needs of small experiments has been developed. 
This paper will discuss not only the development of the gadolinium sulfate compatible resin used in ANNIE, but also the methodology, preparation and test results for several possible resins for use in future detectors.


\section{Measurement Strategy}
\label{sec:strategy}

In a water Cherenkov detector, the two main observables used to assess the quality of the water are light transparency and radiopurity.
Since the latter is not a major concern for the ANNIE experiment, it was not taken into account in the tests described in this paper.
Thus, the ability of the resins, described in Section~\ref{sec:preparation}, to remove uranium from solution has not been measured.
Dedicated studies of gadolinium sulfate radiological purity have been performed by the Super Kamiokande and EGADS collaborations~\cite{Ito:2017zzt}.

The main criterion for success for this application is light transparency, specifically in the 300-500~nm region of the electromagnetic spectrum.
While the Cherenkov spectrum extends to shorter wavelengths than 300~nm, the borosilicate glass used in most conventional photomultiplier tubes (PMT) is opaque to UV light below this regime.
On the other hand, the intensity of Cherenkov light quickly falls at wavelengths longer than 400~nm and becomes practically negligible above 500~nm.
It is worth noting, however, that light absorption outside this region is still important as a tracer of  impurities that, in large amounts, could impact the light transparency in the region of interest.

\begin{figure}[hbt!]
\centering 
\includegraphics[width=0.85\textwidth]{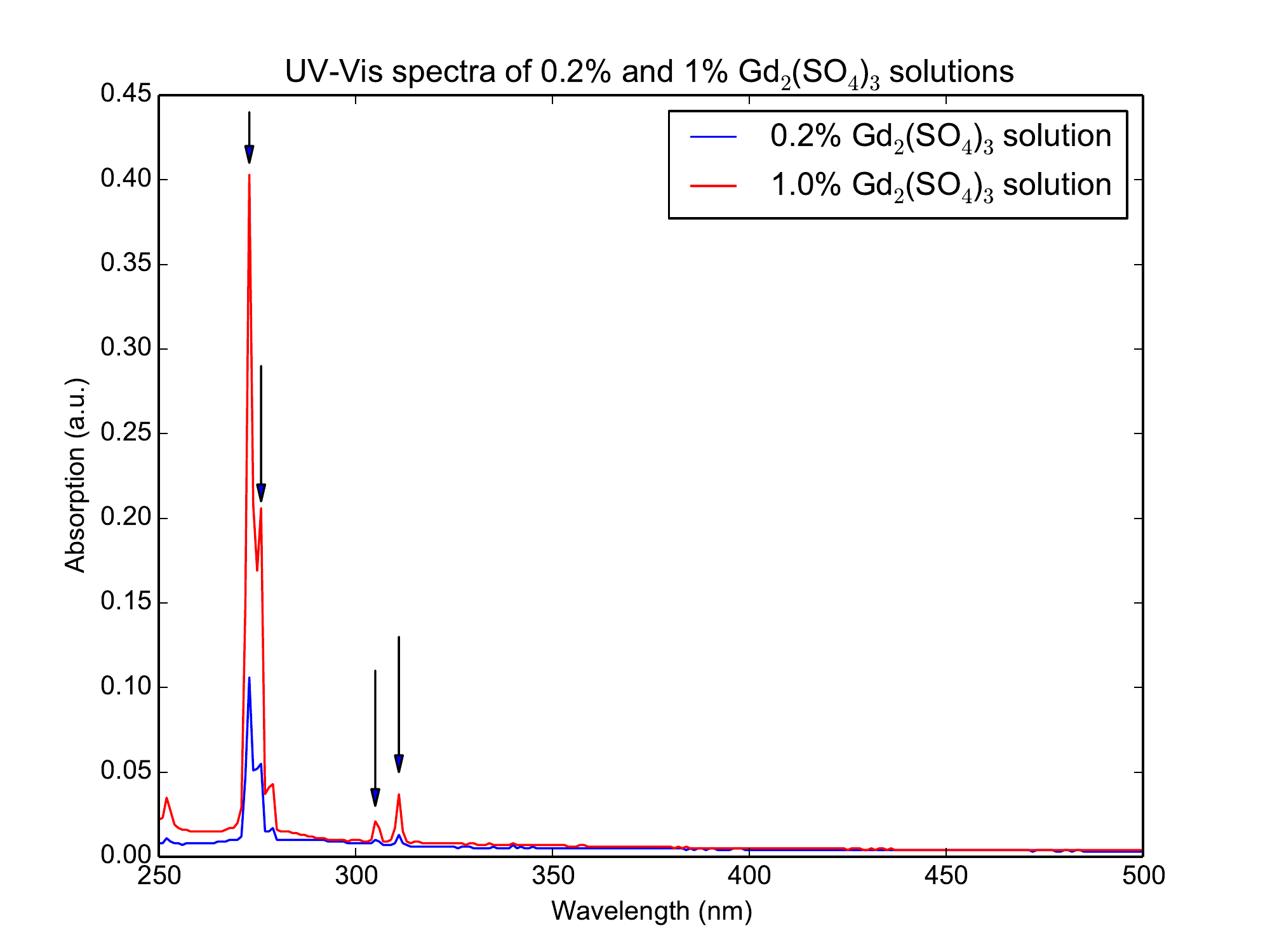}
\caption{UV-Vis spectra of a 1\% w/w (red) and a 0.2\% w/w (blue) \GdSulfate solution. The major absorption peaks observed at 273, 276, 305 and 311~nm are directly related to the concentration of gadolinium, in its ionic form Gd$^{3+}$, in solution.}
\label{fig:UVspectra_pureGdsulfate}
\end{figure}

In addition to maintaining water transparency, the system must be able to monitor and maintain gadolinium sulfate in solution in order to keep the neutron detection efficiency relatively constant. The system must therefore not remove the gadolinium sulfate in any significant quantities while at the same time removing impurities.
For example, the EGADS filtration system is designed for negligible loss of Gd over the course of several years in a 50 kiloton scale detector.
Given the smaller size and the shorter timescale of small experiments like ANNIE (25 tons, which requires roughly 50 kg of gadolinium sulfate), a percent-scale loss of gadolinium over two years is deemed acceptable.

To detect any changes in the concentration of gadolinium in solution, the amplitude of the absorption peaks shown in Figure~\ref{fig:UVspectra_pureGdsulfate} were monitored.
As a cross check the pH of the solution was monitored as well.
The introduction of \GdSulfate in ultrapure DI water tends to lower its pH depending on the concentration of the solute. While the pH of DI water in contact with air usually sits between 4.5 and 6, a 1\% \GdSulfate solution typically has a pH between 4 to 4.5.

Another reason to closely monitor pH is solution stability. While \GdSulfate is soluble in water, other compounds such as gadolinium hydroxide, Gd(OH)$_{3}$, and gadolinium bicarbonate, Gd(HCO$_{3}$)$_{3}$ are not.
The presence of OH$^{-}$ or HCO$_{3}^{-}$ ions, accompanied by an increase in pH, in a Gd-loaded solution can drive Gd$^{3+}$ ions to bond to these ions and drop out of solution as an insoluble precipitate.
Studies showed that the possibility of forming gadolinium hydroxide starts around pH 6.5 and above~\cite{Moeller1951}.
At the other end, a pH approaching a value of 3 is associated with higher acidity of the water and an increased corrosion rate on surrounding materials.
Thus, in the following tests, care is taken to monitor the pH of the gadolinium solutions and to keep it as close as possible to its original value between 4 and 5.
The sensor used to monitor the pH of the solutions was a HANNA~HI~11310 pH electrode calibrated with pH 4.01, 7.01 and 10.01 buffer solutions.

\begin{figure}[hbt!]
\centering 
\includegraphics[width=0.85\textwidth]{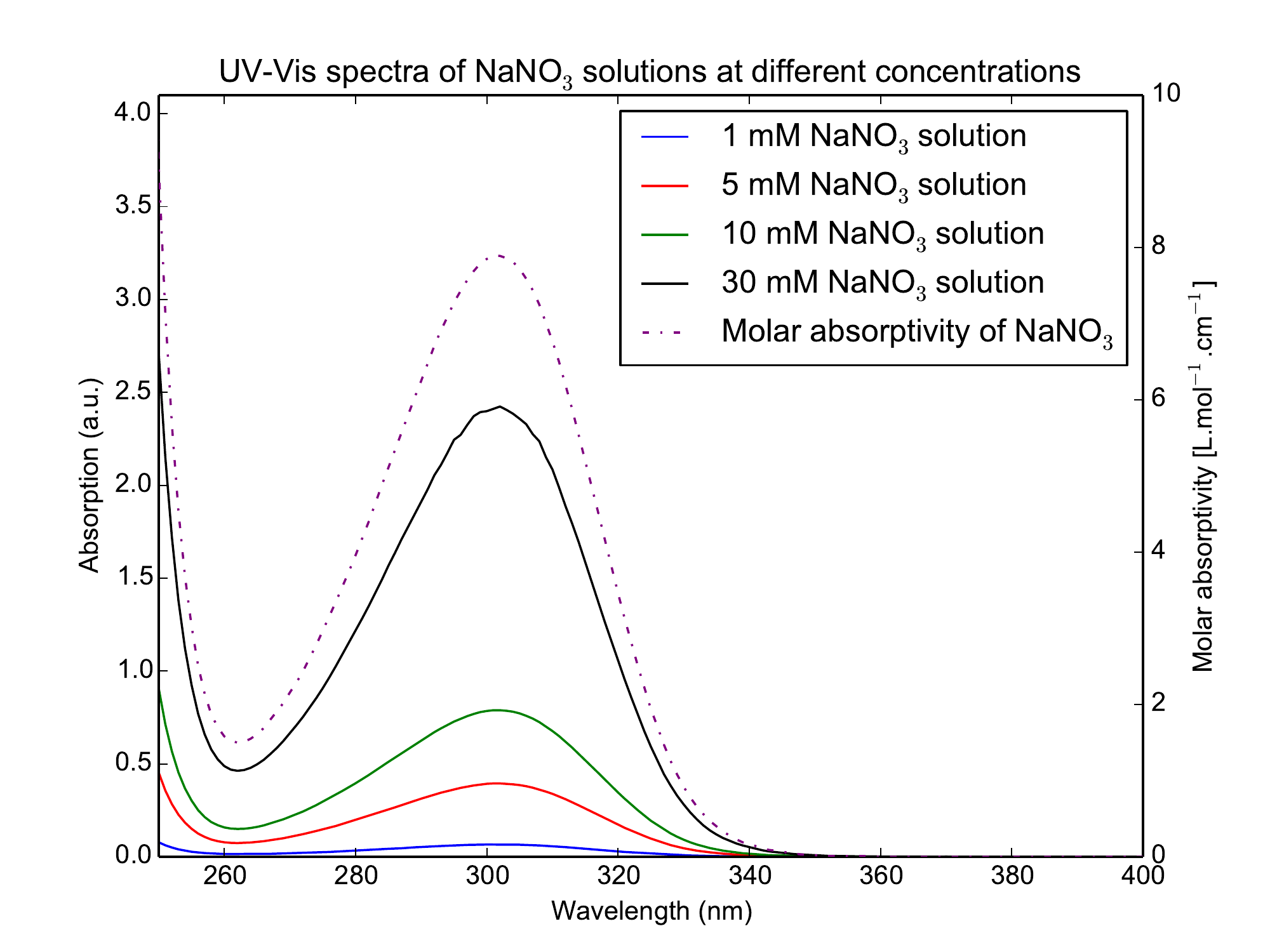}
\caption{UV-Vis spectrum of a 1~millimolar (blue), 5~millimolar (red), 10~millimolar (green) and 30~millimolar (black) sodium nitrate (NaNO$_{3}$) solution. The large absorption is due to the presence of NO$_{3}^{-}$ in solution. The molar absorptivity of sodium nitrate, expressed in L.mol$^{-1}$.cm${-1}$, is displayed as the purple dash dotted line.}
\label{fig:spectra_NaN03}
\end{figure}

Nitrates are a known potential contaminant in gadolinium sulfate due to the use of nitric acid in the gadolinium extraction process.
The concentration of nitrates in the final \GdSulfate solution must be kept to a minimum given the strong absorption in the UV range, as shown in Figure~\ref{fig:spectra_NaN03}.
In addition, leeching of organic UV light absorbing molecules from plastics is also a source of concern. In breaking bonds of organic molecules in the water, water system TOC-UV lamps can also generate free radicals - molecules that are capable of oxidizing other organic and inorganic molecules~\cite{MilliporeTOC}. 
Upon recombination, these radicals can create compounds with a non-negligible UV absorption.
In most applications, these compounds, mostly ionic, are removed from solution with the use of a mixed bed (cation and anion) resin in series with a TOC-UV lamp. Since we do not want to remove gadolinium from solution, only an anion resin is used following the TOC-UV lamp in our system. This is a weakness of our technique, but as will be shown below, even this limited removal system is sufficient for small detectors.

\section{Experimental Setup}
\label{sec:setup}


\subsection{Water filtration system}
\label{subsec:filtration_system}

\begin{figure}[hbt!]
\centering 
\includegraphics[width=0.95\textwidth]{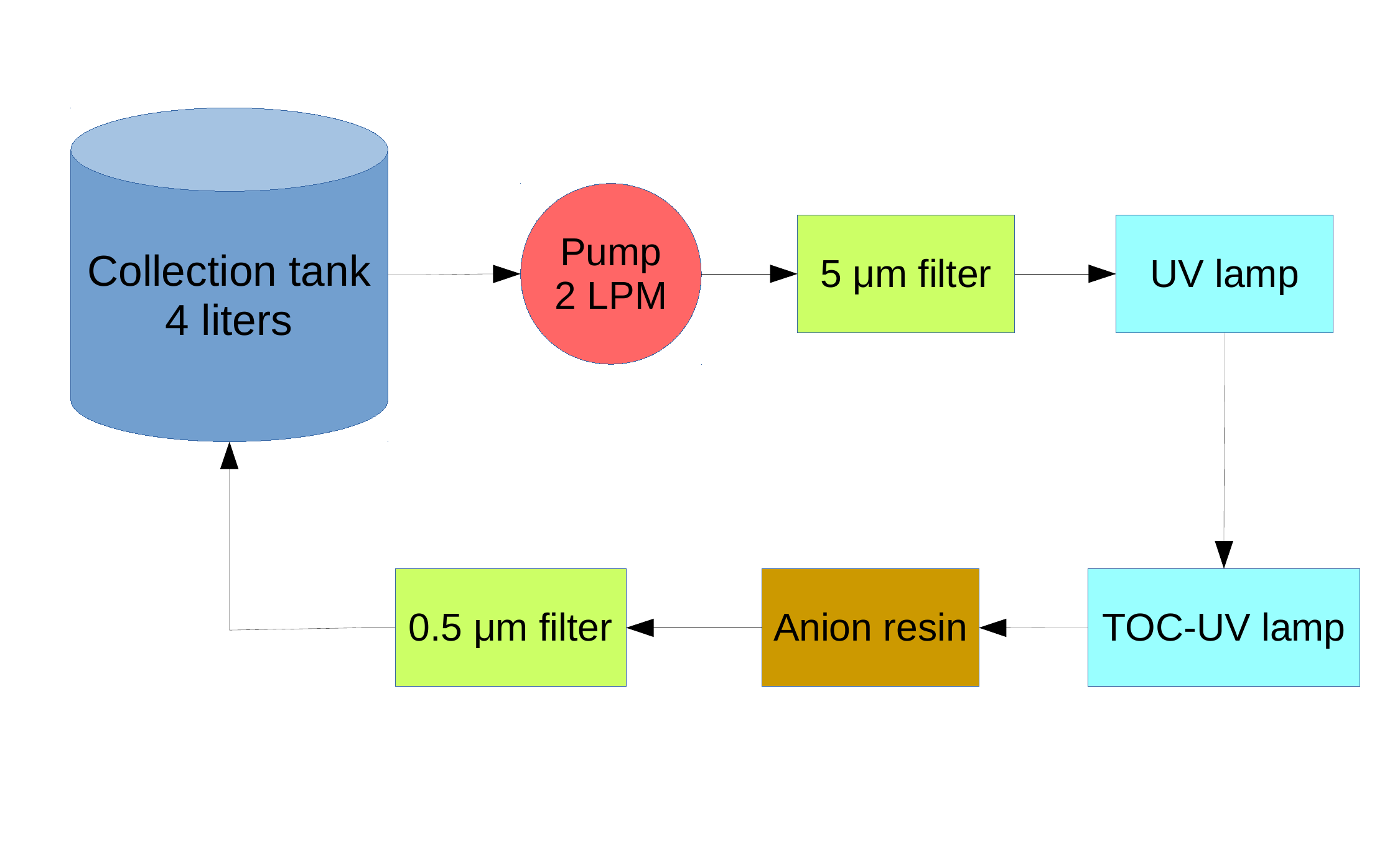}
\caption{Flow diagram of the filtration system used to perform the resin tests. The pump flow and collection tank size allowed a complete circulation of the water in two minutes.}
\label{fig:minisystem_diagram}
\end{figure}

A schematic diagram of the tabletop system used for resin R\&D is shown in Figure~\ref{fig:minisystem_diagram}.
It includes the basic components that will be used to purify Gd-loaded water: two micro-filters (5 and 0.5~$\mu$m rating) for large particulates like dust, a sterilizing UV lamp (254~nm) to limit bacterial growth, a Total Organic Carbon (TOC) UV lamp (185~nm) to break the organic bonds of dissolved plastics, and a canister of anion resins as described in Section~\ref{subsec:resins}.
The system circulates water in a loop from a 4-liter glass container with the use of a diaphragm pump whose flow rate is 2~liters per minute.
All materials used in the system and in contact with water were carefully screened and selected to ensure chemical compatibility with DI and Gd-loaded water.

\subsection{UV-Vis Spectrophotometer}
\label{subsec:spectrometer}

All measurements of light absorption were performed with a Shimadzu UV-1800 UV-Vis spectrophotometer.
A 10-cm quartz cuvette was used, allowing for accurate and consistent measurements of the light attenuation over the range of wavelengths from 200 to 700~nm.
All values and spectra shown hereafter were obtained using ultrapure DI water with a resistivity of 18.2~M$\Omega$.cm as a subtracted baseline.

\subsection{Ion exchange resins}
\label{subsec:resins}

An ion exchange resin is a polymer compound in the form of small plastic beads to be deployed in a flow-through bed.
It has the ability to remove contaminants from a solution and exchange them with alternative compounds embedded in its chemical structure.
Most ion exchange resins have functional groups located at sites within the entire volume of their macroporous beads.
These functional groups attract ions of the opposite charge.
For instance, styrene-based resin beads loaded with trimethylamine functional groups are susceptible to attracting negative ions, hence classifying them as anion resins.
Likewise, cation resins are responsive to positive ions - typically sodium (Na$^{+}$), copper (Cu$^{2+}$) or heavy metals such as lead (Pb$^{2+}$).
In this study, only anion resins will be considered as cation resins have the undesirable effect of removing Gd$^{3+}$ from solution.
Most anion resins used in water treatment applications come preloaded with chloride (Cl$^{-}$) or hydroxide (OH$^{-}$) ions.
The selectivity, or ionic affinity, of such resins causes them to exchange their embedded ions for anions in the water feed such as nitrates (NO$_{3}^{-}$), perchlorates (ClO$_{4}^{-}$) or cyanides (CN$^{-}$). These replacement ions can be as bad for water UV transparency as the contaminants they are replacing, so a new resin must be developed that avoids such compounds. 



\section{Resin preparation}
\label{sec:preparation}

For this study, the three forms of anion resins considered were the chloride (Cl$^{-}$), hydroxide (OH$^{-}$) and sulfate (SO$_{4}^{2-}$).
The use of a Cl$^{-}$ resin with a \GdSulfate solution would lead to the exchange of sulfate in solution for chlorine.
This is not a major flaw in itself, since GdCl$_{3}$ has been used as a compound in detectors in the past~\cite{Dazeley:2015uyd}. However the increasing amounts of chlorine in solution could lead to an accelerated and undesirable corrosion rate~\cite{Coleman:2008ng}.
For this reason, the efforts were focused on the OH$^{-}$ and SO$_{4}^{2-}$ resins.


\subsection{Hydroxide form resin}
\label{subsec:OH_resin}

As mentioned in Section~\ref{sec:strategy}, Gd precipitates in the presence of OH$^{-}$ and becomes the insoluble compound Gd(OH)$_{3}$.
This proscribes the use of a OH$^{-}$ resin directly to purify a \GdSulfate solution.
However, several resins in the hydroxide form have a high selectivity for nitrates over sulfates which is a desirable feature since, as explained in Section~\ref{sec:procedure_results}, nitrates can be a significant contaminant.
In order to take advantage of this high affinity for nitrates while preventing it from releasing OH$^{-}$ ions in the water, the resin must be converted to another form, preferably SO$_{4}^{2-}$ to make it impervious to sulfate ions.
This procedure is similar to the regeneration process performed to rejuvenate resins with their original ions: by flushing them with a concentrated solution of the desired ion.
The \Hydroxide resin chosen for this study was the Purolite A520E due to its high selectivity for nitrates\footnote{The A520E in \Hydroxide form is no longer available as of writing this paper and has been replaced by the A300E in the Purolite catalog.}.\\

In order to convert the A520E resin to a \Sulfate form, it was first flushed with DI water to remove any soluble impurities. The flushed resin was then added to a container of 1~molar (1M) solution of sodium sulfate acidified to a pH of~3 using sulfuric acid.
Immediately after addition, the resin released \Hydroxide ions into solution, raising the pH to above 11. To neutralize these hydroxide ions, additional sulfuric acid was added to the sodium sulfate solution until the pH of the system reached its original value of~3 - a sign that no more \Hydroxide ions were being released.
At this point in the process, the resin was considered to be fully loaded with \Sulfate ions.
However, at this high level of acidity, the concentration of bisulfate ions (HSO$_{4}^{-}$) is significant, and these ions compete with \Sulfate for functional sites on the beads~\cite{BisulfatePaper}.
In order to replace those HSO$_{4}^{-}$ ions mixed in the resin with \Sulfate ions, an alkaline rinse was performed, as described in more detail in  Section~\ref{subsec:SO4_resin}.
Following this process, and a final rinse with DI water, the pH of the resin was measured to be approximately~4, which is deemed acceptable, given the natural pH of a \GdSulfate solution.
This resin, now in a \Sulfate form instead of the \Hydroxide form, will be referred to as the 'prepared \Hydroxide resin' for the remainder of this paper.


\subsection{Sulfate form resin}
\label{subsec:SO4_resin}

The use of an anion resin already in \Sulfate form makes the procedure described in Section~\ref{subsec:OH_resin} unnecessary, however, commercial \Sulfate resins are less common and typically have a lower selectivity for nitrates.
The \Sulfate form resin picked for this study was the Purolite Supergel\texttrademark~SGA550.\\

While a regeneration of the resin in an acidic sulfate-rich bath was not needed, the SGA550 resin exhibited the same behavior as the \Hydroxide resin after regeneration: a significant acidity due to the presence of bisulfate ions.
As the pH of the resin reached a value of about~2.4, an alkaline rinse was again used to neutralize.
This rinse method consisted of putting the resin in a sulfate-rich bath and adding a basic solution of sodium hydroxide (NaOH) in small quantities to neutralize the bisulfate ions as they are released into the solution and replaced by \Sulfate ions.
After performing this alkaline rinse, the pH of the resin was measured to be close to~4, a value comparable to the prepared \Hydroxide form resin.\\

At this point, we now have 2~forms of resins to conduct our studies with, as described below.


\section{Testing procedure and results}
\label{sec:procedure_results}

\subsection{Circulation with a \GdSulfate solution}
\label{subsec:pureGd_circ}

\begin{figure}[hbt!]
\centering 
\includegraphics[width=0.85\textwidth]{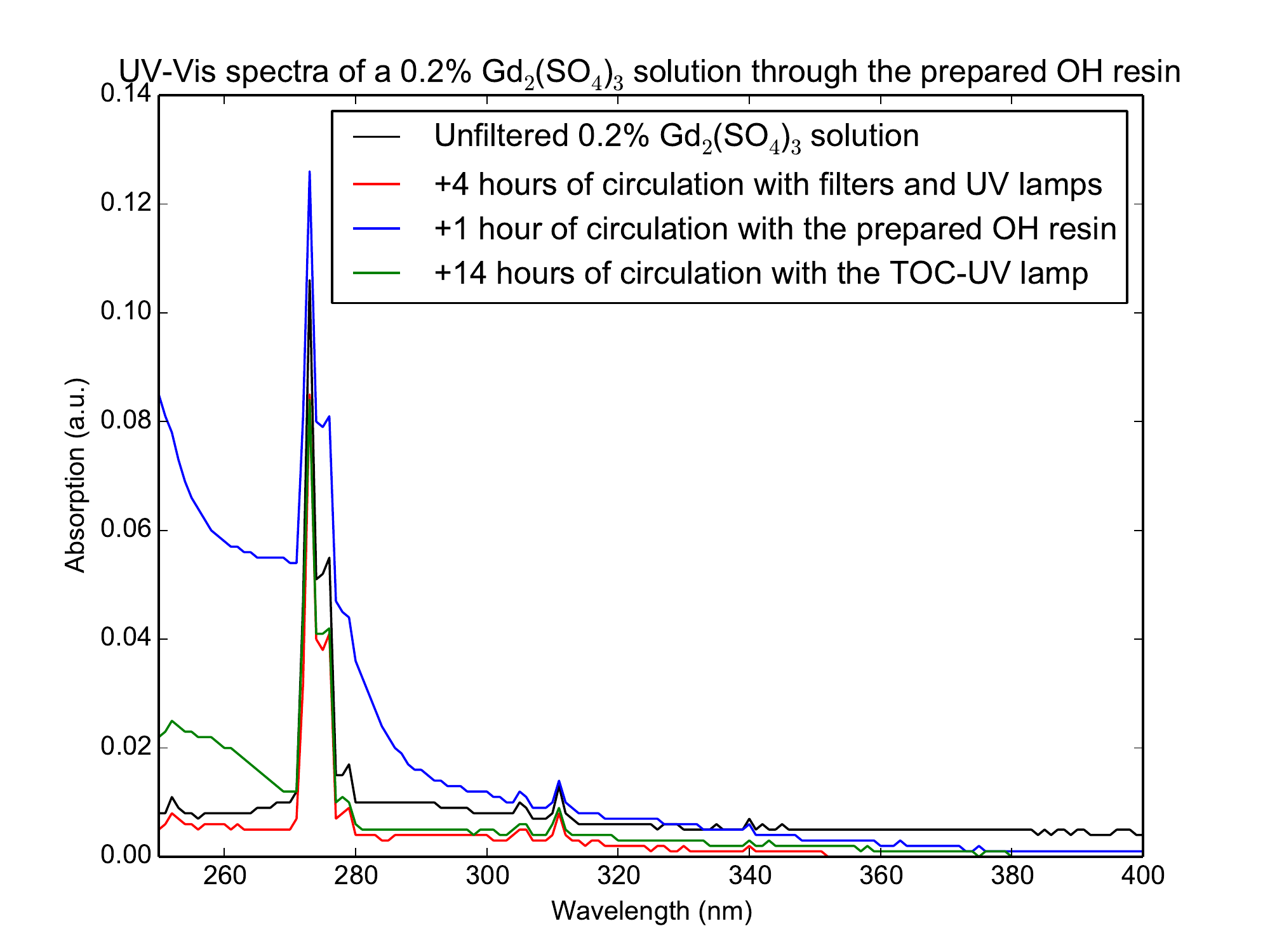}
\caption{UV-Vis spectra of the \GdSulfate stock solution at different steps of the test: Initial solution (grey), stock solution after 4~hours of circulation with filters and UV lamp only (red), after 1~hour of circulation with the prepared \Hydroxide resin added to the system (blue) and after 14~hours of circulation with the TOC-UV lamp on (green).}
  \label{fig:spectra_OHform}
\end{figure}

\begin{figure}[hbt!]
\centering 
\includegraphics[width=0.85\textwidth]{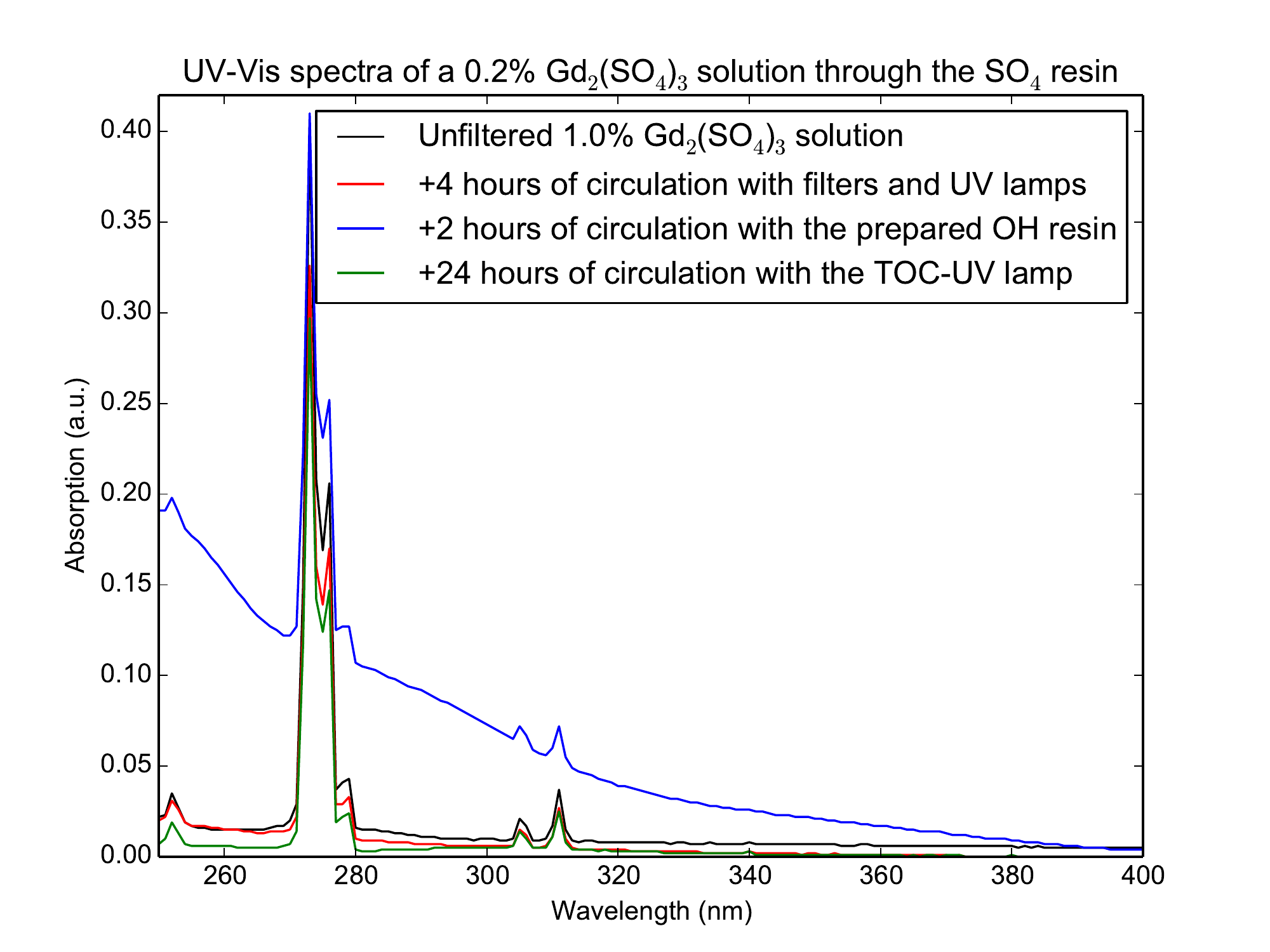}
\caption{UV-Vis spectra of the \GdSulfate solution at different steps of the test: Initial stock solution (grey), after 4~hours of circulation with filters and UV lamp only (red), after 2~hours of circulation with the \Sulfate resin added to the system (blue) and after 24~hours of circulation with the TOC-UV lamp on (green).}
  \label{fig:spectra_SO4form}
\end{figure}

The actions of the prepared \Hydroxide and \Sulfate form resins were tested on two \GdSulfate solutions with concentrations of 0.2\% and 1\% by mass.
Aside from a higher amplitude of the Gd absorption lines in the 1\% solution, as shown in Figure~\ref{fig:UVspectra_pureGdsulfate}, no noticeable differences were observed between the two unfiltered stock solutions.
Prior to introducing the resins into the filtration system, the Gd-loaded water was recirculated through the 2 filters and the UV lamp in order to establish an equilibrium in transparency.
As expected, such pre-filtration increased the water transparency as it removed particulates present in the initial supply of gadolinium sulfate.
The UV-Vis spectra of those unfiltered solutions is shown for reference by the black curves in Figures~\ref{fig:spectra_OHform} and~\ref{fig:spectra_SO4form}.
However, given that the solutions were slightly diluted upon being circulated in the purification system by residual amounts of water left in the filters, they will not be used to assess the capacity of the resins to retain gadolinium in solution.
Once the transparency reached an optimal and stable value, shown by the red curves in the aforementioned figures and taken as a reference hereafter, the test resin was added to the system.

Within minutes, an increase of light absorption was observed, especially at wavelengths shorter than 300~nm, as shown by the blue curves in Figures~\ref{fig:spectra_OHform} and~\ref{fig:spectra_SO4form}.
This broad band absorption was found to be related to the presence of organic molecules, or TOC, in the water.
These compounds are released in solution through leaching and gradual deterioration of the resin.
Ionized or non-ionized, they mostly consist of trimethylamine from the resin's functional groups and other polymer residues~\cite{ResinLeach,TOC_Meyers,TOC_DeSilva}.
This organic leaching effect is the reason why, though the resin is needed to capture free radicals created by the TOC-UV lamp, the TOC-UV lamp is needed to break these organic compounds into free radicals in return.
Both modules must be used in combination to ensure a good optical clarity of the water and care must be taken to ensure that the processes generating organic compounds in the resin have a rate lower than those of the TOC-UV lamp breaking the plastic compounds.

Upon turning on the TOC-UV lamp, the concentration of organic compounds was observed to decrease along with the absorption features they were responsible for, as shown by the green curves in Figures~\ref{fig:spectra_OHform} and~\ref{fig:spectra_SO4form}.
Gd-loaded water was circulated in the system until its transparency reached a stable equilibrium, comparable or better than its original value.
This equilibrium was considered to be reached and stable after 14~hours and 24~hours of recirculation - or 420~and 720~complete turnovers with a flow rate of 2~liters per minute - for the prepared \Hydroxide and \Sulfate form resin, respectively.

The results of these tests are shown in Figure~\ref{fig:spectra_OHform} and Table~\ref{tab:abs_Gd_OHform} for the prepared \Hydroxide form resin and in Figure~\ref{fig:spectra_SO4form} and Table~\ref{tab:abs_Gd_SO4form} for the \Sulfate form resin.
In Tables~\ref{tab:abs_Gd_OHform} and~\ref{tab:abs_Gd_SO4form}, A$_{\text{300~nm}}$ is the value of absorption at 300~nm, A$^{Cher}_{5~m}$ the extrapolated amount of Cherenkov light left after travelling for 5 meters in the solution, and [Gd] the concentration of gadolinium (Gd$^{3+}$) obtained by comparing the amplitudes of the~273 and~276~nm absorption peaks for each spectrum (see Figure~\ref{fig:UVspectra_pureGdsulfate}).
The errors on the absorption values are estimated from the UV-Vis spectrophotometer uncertainty of 0.001 (absorption units).

\begin{table}[hbt!]
\caption{Absorption at 300~nm, intensity of Cherenkov light left after 5~m of the solution and normalized concentration of gadolinium in solution at different steps of the recirculation with the prepared \Hydroxide resin.}
\begin{center}
\begin{tabular}{cccc} \hline
  Step & A$_{\text{300~nm}}$ (a.u.) & A$^{Cher}_{5~m}$ (\%) & [Gd] (\%) \\ 
  \hline
  Filters+UV & 0.004 $\pm$ 0.001 & 88 & 100.0 $\pm$ 2.5 \\
  Filters+UV+Resin & 0.012 $\pm$ 0.001 & 78 & 100.6 $\pm$ 12.3 \\ 
  Filters+UV+Resin+TOC-UV & 0.005 $\pm$ 0.001 & 86 & 93.2 $\pm$ 5.3 \\
\hline
\end{tabular}
\end{center}
\label{tab:abs_Gd_OHform}
\end{table}

\begin{table}[hbt!]
\caption{Absorption at 300~nm, intensity of Cherenkov light left after 5~m of the solution and normalized concentration of gadolinium in solution at different steps of the recirculation with the \Sulfate resin.}
\begin{center}
\begin{tabular}{cccc} \hline
  Step & A$_{\text{300~nm}}$ (a.u.) & A$^{Cher}_{5~m}$ (\%) & [Gd] (\%) \\ 
  \hline
  Filters+UV & 0.006 $\pm$ 0.001 & 90 & 100.0 $\pm$ 1.3 \\
  Filters+UV+Resin & 0.073 $\pm$ 0.001 & 61 & 94.0 $\pm$ 3.0 \\ 
  Filters+UV+Resin+TOC-UV & 0.005 $\pm$ 0.001 & 91 & 93.2 $\pm$ 0.7 \\
\hline
\end{tabular}
\end{center}
\label{tab:abs_Gd_SO4form}
\end{table}

The error associated with the gadolinium concentration is calculated using the instrument's uncertainty, previously mentioned, as well as the deviation between the absorption values before and after the absorption peaks at~273 and~276~nm.
For this reason, the error associated with the gadolinium concentration for spectra with a higher absorption, such as the ones shown by the blue curves in Figures~\ref{fig:spectra_OHform} and~\ref{fig:spectra_SO4form}, is higher.
After introducing the prepared \Hydroxide and the \Sulfate form resins in the system, the gadolinium concentration remained stable within its relative error.
These tests show that the combination of an anion resin and a TOC-UV lamp is capable of purifying Gd-loaded water for hundreds of cycles without introducing light-absorbing contaminants into solution or significantly affecting the gadolinium concentration in the water.
While the two resins showed similar satisfactory behavior, the \Sulfate form requires a more straightforward preparation and is commercially available.
The ANNIE collaboration thus decided to use it in the experiment's water filtration system.
With a complete turnover of the water volume every $\sim$55 hours, this resin is not expected to significantly affect the gadolinium concentration in the ANNIE water even after 4~years of operation.
Henceforth, all subsequent tests will be performed with this resin.

From the literature, it is assumed that the contribution of the anion resins to the overall transparency of the water is made by the capture of free radicals from organic molecules being broken by the TOC-UV lamp. 
As the concentration of these radicals downstream of the TOC-UV lamp is inaccessible, it was not possible to definitively assess the efficacy of the resin to remove contaminants. 
Therefore, an external contamination was introduced in the \GdSulfate solution to quantify the removal capabilities of the resin.

\subsection{Tests with an artificially contaminated \GdSulfate solution}
\label{subsec:NO3_circ}

This test was performed using the \Sulfate form resin after ensuring its ability to purify Gd-loaded water for more than 48~hours, as shown in Section~\ref{subsec:pureGd_circ}. 
The contaminant chosen for this test was \Nitrate in the form of sodium nitrate (NaNO$_{3}$) since, as discussed previously, nitrate is a common contaminant in Gd-loaded water.

\begin{figure}[hbt!]
\includegraphics[width=0.85\linewidth]{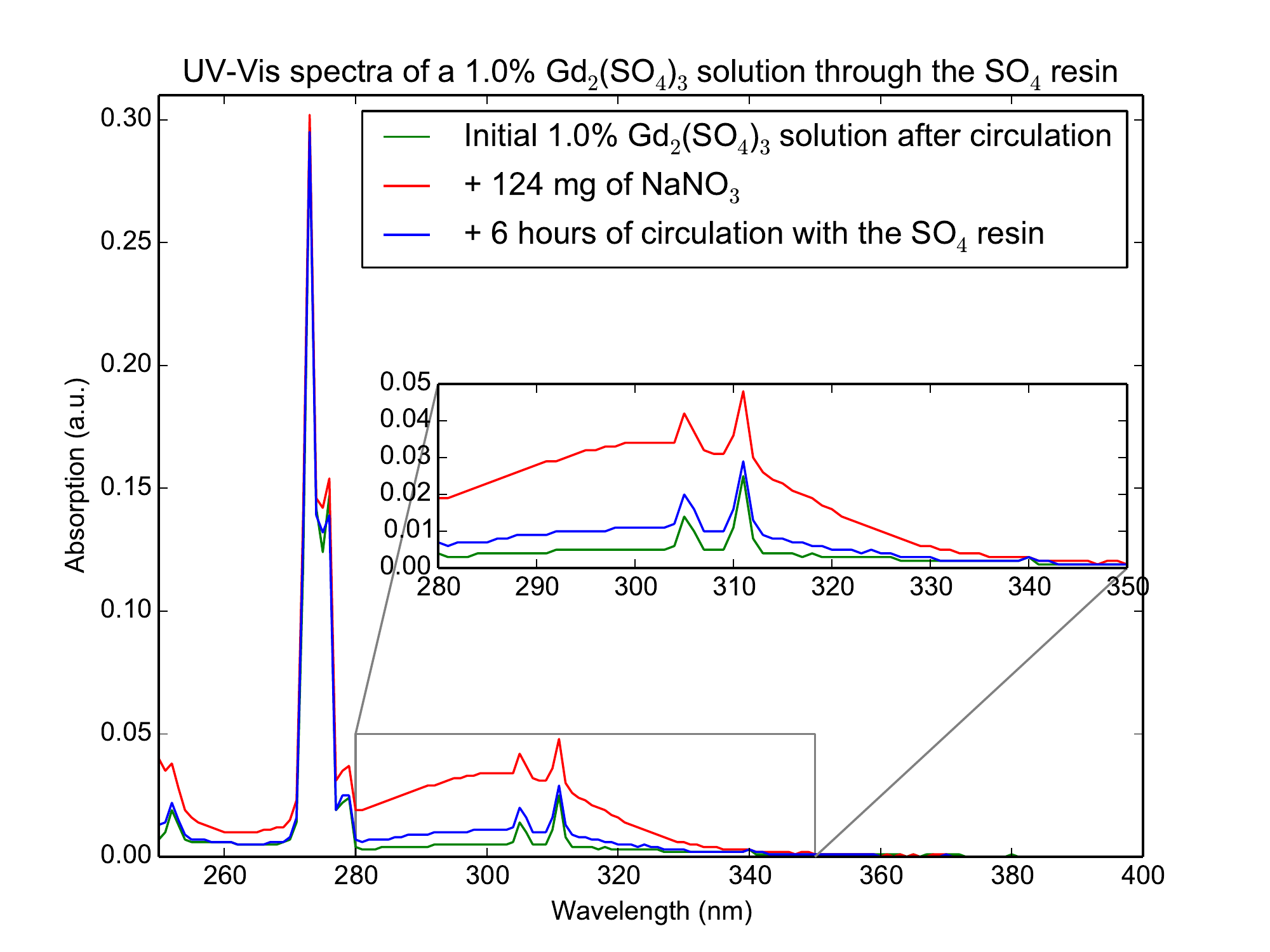}
\centering 
\caption{UV-Vis spectra of the \GdSulfate solution at different steps of the nitrate contamination test using the \Sulfate form resin: Initial solution after more than 48 hours of circulation (grey), after adding 124~mg of nitrate (dashed red) and after 6~hours of circulation (blue).}
\label{fig:spectra_SO4form_nitrate}
\end{figure}

A reduction of the nitrate concentration was observed, demonstrating its removal capabilities. 
This nitrate removal was not accompanied by any significant loss of gadolinium concentration or a change in pH.
As seen in Figure~\ref{fig:spectra_SO4form_nitrate}, the large absorption centered around 300~nm, characteristic of NO$_{3}^{-}$, is largely suppressed after the action of the resin.
By measuring the concentration of nitrates in the solution, directly related to the amplitude of the nitrate UV absorption, before and after passing it through the resin, it was determined that 100~grams of the commercial \Sulfate form resin removed 100~mg of \Nitrate ions.

This test shows that a kilogram of \Sulfate resin is capable of removing at least 1~gram of contaminants in the form of \Nitrate ions.


\section{Summary}
\label{sec:summary}

The goal of these tests was to validate the use of ion exchange resins in the \Sulfate form (out-of-the-box or prepared) as a way to purify Gd-loaded water and describe how one could prepare such resins.
Resins capable of capturing anionic contaminants, such as nitrates and free radicals, can be developed at a low price and without intricate chemical processes using commercially-available anion resins.
While additional tests are needed to assess all purifying capabilities of these resins, no significant loss of gadolinium or light transparency in the UV region of interest for water Cherenkov detectors was observed after numerous recirculation and purification cycles.
Although not investigated in this paper, anion resins may have the capability to remove uranium.
The anion resin mentioned in Ref.~\cite{Ikeda:2019pcm} was initially used to remove uranium from Gd-loaded water since, in a sulfate-rich environment, uranium sulfate complexes (U(SO$_{4}$)$_{2}$) are prone to form.
Anion resins with a high selectivity for sulfate ions could thus be used to remove uranium from solutions.
This may be a fruitful topic for future research.
In the context of neutrino and low backgrounds physics, the quantity of uranium contaminants would be infinitesimally smaller but the capability to reduce their concentration in Gd-loaded would have a major impact on low energy event detection.
The development of such anion resins is expected to have a substantial significance on the design, capabilities and cost of water purification systems for future experiments using water loaded with gadolinium sulfate as their detection medium.\\

\acknowledgments

This work was supported by the U.S. Department of Energy (DOE) Office of Science under award number DE-SC0009999, and by the DOE National Nuclear Security Administration through the Nuclear Science and Security Consortium under award number DE-NA0003180. 
The authors would like to thank UC Davis students Lena Korkeila and Amilcar Perez for their assistance throughout the tests.
The authors particularly wish to thank Mark Vagins for useful discussions and guidance regarding gadolinium sulfate and water filtration systems, and the ANNIE Collaboration in general for motivating this work.


\end{document}